\documentclass[]{article}
\usepackage{url}
\usepackage[]{amsmath}
\usepackage[]{amssymb}
\usepackage[]{amsthm}
\usepackage{geometry}
\usepackage{rotating}
\usepackage{enumitem}
\usepackage{natbib}
\usepackage[onehalfspacing]{setspace}
\usepackage{graphicx}
\usepackage{tikz}
\usepackage{tabu}
\usetikzlibrary{positioning}

\def\Cov {{ \mathrm{cov} }}

\def\se {{ \mathrm{se} }}

\DeclareMathOperator*{\argmin}{arg\,min}

\begin{document}

% Title of paper
\title{An efficient and robust approach to Mendelian randomization with measured pleiotropic effects in a high-dimensional setting}

% List of authors, with corresponding author marked by asterisk
\author{Andrew J. Grant$^{1}$\footnote{Corresponding author. Email address: andrew.grant@mrc-bsu.cam.ac.uk}, Stephen Burgess$^{1, 2}$\\[6pt]
\textit{$^1$MRC Biostatistics Unit, University of Cambridge, Cambridge, UK}\\
\textit{$^2$Cardiovascular Epidemiology Unit, University of Cambridge, Cambridge, UK}}
\date{}

\maketitle

\begin{abstract}
Valid estimation of a causal effect using instrumental variables requires that all of the instruments are independent of the outcome conditional on the risk factor of interest and any confounders. In Mendelian randomization studies with large numbers of genetic variants used as instruments, it is unlikely that this condition will be met. Any given genetic variant could be associated with a large number of traits, all of which represent potential pathways to the outcome which bypass the risk factor of interest. Such pleiotropy can be accounted for using standard multivariable Mendelian randomization with all possible pleiotropic traits included as covariates. However, the estimator obtained in this way will be inefficient if some of the covariates do not truly sit on pleiotropic pathways to the outcome. We present a method which uses regularization to identify which out of a set of potential covariates need to be accounted for in a Mendelian randomization analysis in order to produce an efficient and robust estimator of a causal effect. The method can be used in the case where individual-level data are not available and the analysis must rely on summary-level data only. It can also be used in the case where there are more covariates under consideration than instruments, which is not possible using standard multivariable Mendelian randomization. We show the results of simulation studies which demonstrate the performance of the proposed regularization method in realistic settings. We also illustrate the method in an applied example which looks at the causal effect of urate plasma concentration on coronary heart disease.
\end{abstract}

\section{Introduction}

Instrumental variables can be used to estimate the causal effect of an exposure (also called a risk factor) on an outcome from observational data. A variable is a valid instrument if it is: associated with the risk factor; independent of any confounders of the association between the risk factor and the outcome; and independent of the outcome conditional on the risk factor and confounders. These are the three instrumental variables assumptions \citep{Greenland2000}.

\begin{figure}
	\centering
	\begin{tikzpicture}[]
	\node[] (X) {$X$};
	\node[] (G1) [left=3cm of X] {$G_{1}$};
	\node[] (Gdots) [below=0.5cm of G1] {$\vdots$};
	\node[] (Gp) [below=0.5cm of Gdots] {$G_{p}$};
	\node[] (W1) [below=of X] {$W_{1}$};
	\node[] (Wdots) [below=0.5cm of W1] {$\vdots$};
	\node[] (Wk) [below=0.5cm of Wdots] {$W_{k}$};
	\node[] (U) [above right=2.24cm of X] {$U$};
	\node[] (Y) [right=3cm of X] {$Y$};
	
	\path[-stealth]
	(G1) edge (X)
	(G1) edge (W1)
	(G1) edge (Wk)
	(Gp) edge (X)
	(Gp) edge (W1)
	(Gp) edge (Wk)
	(W1) edge (Y)
	(Wk) edge (Y)
	(X) edge (Y)
	(U) edge (X)
	(U) edge (Y)
	(U) edge (W1)
	(U) edge (Wk);
	\end{tikzpicture}
	\caption{Directed acyclic graph showing the associations between the genetic variants ($G_{1}, \ldots, G_{p}$), the risk factor ($X$), measured covariates which potentially give rise to pleiotropy ($W_{1}, \ldots, W_{k}$), potentially unknown and unmeasured confounders ($U$) and the outcome ($Y$).}
	\label{fg:dag}
\end{figure}
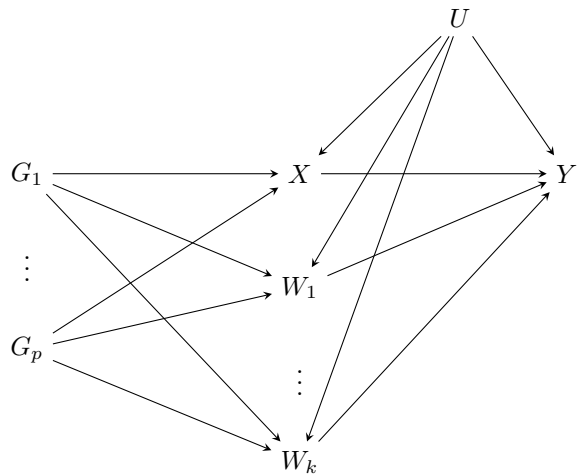

In Mendelian randomization studies, genetic variants are used as instrumental variables \citep{Katan1986,GDS2003,Lawlor2008}. Although genetic variants have many properties which make them attractive candidates for instruments, one disadvantage is that a single variant typically explains only a small amount of the variation in a risk factor. It is therefore advantageous to combine information from a number of genetic variants. Given the proliferation of genome-wide association studies (GWAS) in recent years, there is data available linking genetic variants across the entire human genome to an enormous number of traits. Standard instrumental variables and meta-analysis techniques allow us to combine the individual estimates given by each of these genetic variants \citep{ThompsonSharp1999,Palmer2012}. However, the more genetic variants that are added to an analysis, the more likely that at least one of them will be an invalid instrument. In particular, any given genetic variant could associate with a number of traits other than the risk factor of interest. If any of these traits, which we refer to as covariates, associate with the outcome via pathways which bypass the risk factor, then the third instrumental variables assumption is violated and estimates of the causal effect will be biased. This is known as pleiotropy. This scenario is illustrated via the directed acyclic graph in Figure \ref{fg:dag}.

There are many methods for estimating a causal effect in the presence of pleiotropy. However, such methods typically require at least some of the genetic variants to be valid instruments. These include median-based estimators \citep{Bowden2016median}, mode-based estimators \citep{Hartwig2017mode,Guo2018} and the contamination mixture method \citep{BF2019contamination}. The MR-Egger method \citep{Bowden2015egger} consistently estimates the causal effect without requiring the assumption of no pleiotropy. However, the method relies on a different assumption that pleiotropic effects are independent of the genetic variant-risk factor associations. This assumption is almost as strong as the one it replaces and hypothesis testing based on the MR-Egger method often has low power. Regularization methods proposed by \cite{Kang2016}, \cite{Windmeijer2018} and \cite{Rees2019} use $\ell_{1}$ penalization of the least squares equation to down-weight, and possibly remove, invalid instruments. These methods implicitly assume that at least some of the instruments are valid, and \cite{Kang2016} show that consistent estimation requires a majority of instruments to be valid.

\cite{Jiang2019} proposed a constrained optimization approach to construct a weighting scheme for the genetic variants that balances the pleiotropic effects. The weighting scheme can be thought of as weights used to construct an allele score \citep{BThompson2013allelescore, BDudbridge2016allele}, which can then be used as an instrument in place of the genetic variants themselves. In Section \ref{se:covbalancing} we demonstrate that, in the case where the number of genetic variants is greater than the number of covariates, the estimator obtained in this way is equivalent to that obtained using standard multivariable Mendelian randomization \citep{BThompson2015mv}. The interpretation of this is that we can account for covariates which give rise to pleiotropy by including them in a multivariable Mendelian randomization model.

In practice, it will very often be the case that only a relatively small number of potential covariates need to be included in a multivariable analysis in order to balance pleiotropy. That is, only some of the traits will actually sit on pathways to the outcome which bypass the risk factor. If this is the case, then the estimator of the causal effect obtained by a multivariable Mendelian randomization analysis with all potential covariates included will be inefficient. Furthermore, if there are more covariates than genetic variants, multivariable Mendelian randomization cannot be performed, since this is akin to fitting a regression model with more predictor variables than observations.

In this paper we propose a method for estimating a causal effect where any number of the instruments are invalid due to measured pleiotropy. The method identifies the covariates, among a set of potential covariates, which are on causal pathways from the genetic variants to the outcome, and which therefore should be accounted for in a multivariable Mendelian randomization analysis. Our approach is to fit a multivariable model which applies an $\ell_{1}$ penalty on the coefficients of the genetic variant-covariate associations without applying penalization on the coefficient of the genetic variant-risk factor association. The coefficients of the genetic variant associations with the covariates which have little or no pleiotropic effects will be shrunk towards zero. We thus obtain a more efficient estimator of the causal effect than we would by controlling for all covariates, and a less biased estimate than we would by ignoring the covariates. There are existing methods which apply regularization to covariates (see, for example, \citealp{Caner2009} and \citealp{Fan2014}). The method of \cite{Lin2015} is in fact a two stage procedure which regularizes both the covariates and the instruments. Our situation differs from these, however, in that we do not wish to apply penalization to the coefficient of the risk factor, but only those of the covariates. That is, not all coefficients in the model are penalized, which is a non-standard scenario. The approach is developed for use with summarized data, that is, when the only data available are estimates of the genetic associations with the risk factor, covariates and outcome, and their standard errors. This is typically the way in which GWAS data are made available. It does not require any of the genetic variants to be free of pleiotropy and, furthermore, since it is based on regularization, the method can be used in the case where there are more covariates under consideration than instruments.

In Section \ref{se:model} we formally define the model under consideration and demonstrate the inconsistency of standard instrumental variables estimators if pleiotropy is ignored. In Section \ref{se:covbalancing} we show how to construct covariate balancing allele scores based on the constrained optimization approach using summarized data, and demonstrate the equivalence of this approach to standard multivariable Mendelian randomization. In Section \ref{se:regmethod} we describe our method for estimating the causal effect by applying regularization to the coefficients of the genetic variant-covariate associations. In Section \ref{se:sims} we present the results of simulation studies which examine the performance of the method. In Section {\ref{se:appliedex} we demonstrate the approach with an applied example which looks at the effect of plasma urate concentration on coronary heart disease.
	
\section{The model} \label{se:model}

For individual $i$, let $Y_{i}$ be the outcome, $X_{i}$ be the risk factor, $G_{i}=\begin{bmatrix} G_{i1} & \cdots & G_{ip} \end{bmatrix}'$ be genetic variants and $W_{i}=\begin{bmatrix} W_{i1} & \cdots & W_{ik} \end{bmatrix}'$ be covariates potentially on the causal pathway between each of the genetic variants and the outcome. The model we consider is given by
\begin{align}
X_{i} &= G_{i}' \beta_{X} + \gamma_{X} U_{i} + \varepsilon_{Xi} \label{eq:X}\\
W_{ij} &= G_{i}' \beta_{Wj} + \gamma_{Wj} U_{i} + \varepsilon_{Wij}, \quad j=1,\ldots,k \label{eq:Wj} \\
Y_{i} &= \theta X_{i} + W_{i}' \delta + \gamma_{Y} U_{i} + \varepsilon_{Yi} \label{eq:Y} ,
\end{align}
where $\beta_{X}$ is a $p \times 1$ vector of regression coefficients representing the associations between the $p$ genetic variants and the risk factor and $\beta_{Wj}$ is a $p \times 1$ matrix of regression coefficients representing the associations between the $p$ genetic variants and the $j$th covariate. The variable $U_{i}$ represents confounders of the associations between the risk factor, covariates and outcome. The parameters $\theta$, $\gamma_{X}$, $\gamma_{W1}, \ldots, \gamma_{Wk}$ and $\gamma_{Y}$ are scalars and $\delta = \begin{bmatrix} \delta_{1} & \cdots & \delta_{k} \end{bmatrix}'$ is a $k \times 1$ vector. It is assumed that the noise terms, $\varepsilon_{Xi}, \varepsilon_{Wi1}, \ldots, \varepsilon_{Wik}$, and $\varepsilon_{Yi}$ are independent of $U_{i}$ and the genetic variants. It is also assumed that the genetic variants are independent of each other (that is, no linkage disequilibrium) and independent of $U_{i}$. By assuming that $U_{i}$ is independent of the $G_{i}$, the second instrumental variables assumption is satisfied and so the only violation of the assumptions is that of pleiotropy.

We let $\hat{\beta}_{Xi}$, $\hat{\beta}_{Wij}$ and $\hat{\beta}_{Yi}$ be the estimates of the associations between the $i$th genetic variant and the risk factor, the $j$th covariate and the outcome, respectively. We denote by $\hat{\beta}_{X}$ and $\hat{\beta}_{Y}$ the $p \times 1$ vectors with $i$th elements $\hat{\beta}_{Xi}$ and $\hat{\beta}_{Yi}$, respectively, and $\hat{\beta}_{W}$ the $p \times k$ matrix with $\left( i,j \right)$th element $\hat{\beta}_{Wij}$. While instrumental variable analyses can be performed using individual-level data, often in practice only summarized data in the form of these regression coefficients and their standard errors are available to investigators. To aid applicability of the method, our method is formulated using these summarized data only.

If the three instrumental variables assumptions are met, the causal effect parameter $\theta$ can be consistently estimated using the two stage least squares method. In the first stage, the risk factor is regressed on the genetic variants. In the second stage, the outcome is regressed on the fitted values from the first stage. The regression estimate from the second stage is the estimate of the causal effect. When only summarized data is available, the same estimator can be obtained by using the inverse-variance weighted method \citep{BButterworthThompson2013}, which fits the regression model
\[
\hat{\beta}_{Yj} = \theta \hat{\beta}_{Xj} + \varepsilon_{j},
\]
where $\varepsilon_{j}$ is assumed to be normally distributed with mean zero and variance equal to the variance of $\hat{\beta}_{Yj}$, denoted $\se^{2} \left( \hat{\beta}_{Yj} \right)$. That is, the inverse-variance weighted estimator is
\[
\hat{\theta}_{IVW} = \frac{\hat{\beta}_{X}' S \hat{\beta}_{Y}}{\hat{\beta}_{X}' S \hat{\beta}_{X}} ,
\]
where $S$ is the $p \times p$ diagonal matrix with  $(j,j)$th element $\se^{-2} \left( \hat{\beta}_{Yj} \right)$. Under the model considered here,
\[
\hat{\theta}_{IVW} \rightarrow_{p} \theta + \frac{\beta_{X}' \Sigma_{G} \beta_{W} \delta}{\beta_{X}' \Sigma_{G} \beta_{X}} ,
\]
where $\rightarrow_{p}$ denotes convergence in probability, $\beta_{W} = \begin{bmatrix} \beta_{W1} & \cdots & \beta_{Wk} \end{bmatrix}$ and $\Sigma_{G}$ is the $p \times p$ matrix with $\left( i,j \right)$th element the covariance of the $i$th and $j$th genetic variant. Thus, $\hat{\theta}_{IVW}$ is an inconsistent estimator of the causal effect if $\beta_{W} \delta \neq 0$, that is, if pleiotropy is present.

\section{Covariate balancing} \label{se:covbalancing}
\cite{Jiang2019} proposed weighting the genetic variants in such a way that the pleiotropic effects are balanced out. Such a weighting scheme, however, will tend to reduce the strength of the association between the risk factor and the weighted genetic variants. A constrained optimization approach was therefore proposed, which aims to maximise the covariance between the weighted genetic variants and the risk factor under the constraint that the covariances between the weighted genetic variants and each of the covariates are zero.

We can adapt the constrained optimization approach to the summarized data case as follows. Letting $\alpha$ be a $p \times 1$ vector of weights, $\Cov \left( G \alpha, X \right) = \alpha' \Sigma_{G} \beta_{X}$ and $\Cov \left( G \alpha, W \right) = \alpha' \Sigma_{G} \beta_{W}$. We thus wish to maximise with respect to $\alpha$ the objective function $\alpha' \Sigma_{G} \hat{\beta}_{X}$, subject to $\alpha' \Sigma_{G} \hat{\beta}_{W} = 0$ and $\alpha' \Sigma_{G} \alpha = 1.$ The second constraint is a normalising condition so that a unique solution is possible. When $p>k$, this can be solved in closed form by
\begin{equation}
\alpha = \tilde{\alpha} = \frac{\xi}{\xi' \Sigma_{G} \xi}, \label{eq:conopt1}
\end{equation}
where
\begin{equation}
\xi = \hat{\beta}_{X} - \hat{\beta}_{W} \left( \hat{\beta}_{W}' \Sigma_{G} \hat{\beta}_{W} \right)^{-1} \left( \hat{\beta}_{W}' \Sigma_{G} \hat{\beta}_{X} \right) . \label{eq:conopt2}
\end{equation}
The causal effect is then estimated by
\begin{equation}
\frac{\tilde{\alpha}' \Sigma_{G} \hat{\beta}_{Y}}{\tilde{\alpha}' \Sigma_{G} \hat{\beta}_{X}} . \label{eq:conopt3}
\end{equation}
In practice, $\Sigma_{G}$ is unknown. However, since $S$ is approximately proportional to $\Sigma_{G}$, we can replace $\Sigma_{G}$ by $S$ in (\ref{eq:conopt1}), (\ref{eq:conopt2}) and (\ref{eq:conopt3}). It is shown in the Appendix that this estimator is the same as that obtained by fitting the weighted linear regression model
\begin{equation}
\hat{\beta}_{Yj} = \theta \hat{\beta}_{Xj} + \delta_{1} \hat{\beta}_{Wj1} + \cdots + \delta_{1} \hat{\beta}_{Wjk} + \varepsilon_{j}, \label{eq:mvmr}
\end{equation}
where $\varepsilon_{j}$ is normally distributed with mean zero and variance $\se^{2} \left( \hat{\beta}_{Yj} \right)$. This is the multivariable inverse-variance weighted method \citep{BDudbridgeThompson2015mv}. Thus, we obtain an estimator of the causal effect which controls for measured pleiotropy by using a standard multivariable Mendelian randomization approach. However, as noted above, this estimator will be inefficient if any of the covariates do not sit on pathways between the genetic variants and the outcome which bypass the risk factor.

\section{The regularization method} \label{se:regmethod}

\subsection{Estimating the causal effect}
Suppose we believe that not all $k$ covariates have pleiotropic effects. That is, that some of the $\delta_{j}$'s are zero. We can induce sparsity in $\delta$ by including an $\ell_{1}$ penalty term in the least squares equation used for estimating the parameters in (\ref{eq:mvmr}). That is, the parameter estimators are given by
\begin{equation}
\argmin_{\theta, \delta} \frac{1}{2}\left( \hat{\beta}_{Y} - \theta \hat{\beta}_{X} -  \hat{\beta}_{W} \delta \right)' S \left( \hat{\beta}_{Y} - \theta \hat{\beta}_{X} -  \hat{\beta}_{W} \delta \right) + \lambda \sum_{i=1}^{k} \left| \delta_{i} \right|, \label{eq:kreg}
\end{equation}
where $\lambda>0$ is a tuning parameter. This is not a standard Lasso problem, since we are not penalizing all the parameters in the model. It is analogous to the some valid, some invalid IV estimator (sisVIVE) of \cite{Kang2016}, which also minimizes a sum of squares function with all but one parameter subject to penalization. In the sisVIVE setup, invalid instruments are identified by applying penalization on direct effects between the instruments and the outcome, but the causal effect is not subject to penalization. Our case is different in that we do not seek to identify valid instruments, but rather to identify pleiotropic covariates using summarized data.

Following a similar procedure to that of the proof of Theorem 3 in \cite{Kang2016}, it is shown in the Appendix that the estimator of $\theta$ obtained by (\ref{eq:kreg}), for a given value of $\lambda$, is equivalent to that given by the following two step procedure.
\begin{enumerate}
	\item Let
	\[
	\hat{\delta}_{\lambda} = \argmin_{\delta} \left( \hat{\beta}_{Y} - \hat{\beta}_{W} \delta \right)' S^{1/2} P_{b^{\bot}} S^{1/2} \left( \hat{\beta}_{Y} - \hat{\beta}_{W} \delta \right) + \lambda \sum_{i=1}^{k} \left| \delta_{i} \right|,
	\]
	where $P_{b^{\bot}} = I_{p} - \hat{\beta}_{X} \left( \hat{\beta}_{X}' S \hat{\beta}_{X} \right)^{-1} \hat{\beta}_{X}'$ .
	\item Let
	\[
	\hat{\theta}_{\lambda} = \frac{\left( \hat{\beta}_{Y} - \hat{\beta}_{W} \hat{\delta}_{\lambda} \right)' S \hat{\beta}_{X}}{\hat{\beta}_{X}' S \hat{\beta}_{X}} .
	\]
\end{enumerate}

The first step is now a standard Lasso problem. It induces shrinkage on the elements of $\delta$, but not on $\theta$. Some of the elements of $\delta$ will be shrunk to zero, and the corresponding covariates are effectively removed from the analysis. The second step can be interpreted as estimating $\theta$ by a weighted regression of $\hat{\beta}_{Y} - \hat{\beta}_{W} \hat{\delta}_{\lambda}$ on $\hat{\beta}_{X}$.

An alternative estimator is obtained by dropping the covariates that are assigned a zero coefficient by the above procedure and then performing a standard multivariable analysis including the remaining covariates. That is, the two step procedure is effectively used as a model selection technique. This is along the lines of, for example, the post-Lasso estimators of \cite{belloni2012} and \cite{Windmeijer2018}, and the LARS-OLS hybrid estimator of \cite{Efron2004}. The main argument for using such post-regularization estimators is that they avoid potential bias that may arise from the shrinkage of some of the regression coefficients. The cost is some loss in efficiency.

\subsection{The choice of tuning parameter}
An important aspect of the method is the choice of tuning parameter, $\lambda$, which controls the level of sparsity. A common approach to choosing $\lambda$ is $K$-fold cross-validation. The set of genetic variants is split into $K$ folds, and the estimation procedure is performed, over a range of $\lambda$ values, holding out each fold in turn. The $\lambda$ chosen is that which minimizes the mean, across each fold, of a particular target function. A natural choice for the cross-validation target function is the mean squared error, that is
\begin{equation}
\frac{1}{p}\left( \hat{\beta}_{Y} - \hat{\beta}_{W} \hat{\delta}_{\lambda} - \hat{\theta}_{\lambda} \hat{\beta}_{X} \right)' S \left( \hat{\beta}_{Y} - \hat{\beta}_{W} \hat{\delta}_{\lambda} - \hat{\theta}_{\lambda} \hat{\beta}_{X} \right) . \label{eq:cv2}
\end{equation}
An alternative is to make the choice of $\lambda$ in Step 1, independent of $\hat{\beta}_{X}$. That is, the cross-validation target function is
\begin{equation}
\frac{1}{p}\left( \hat{\beta}_{Y} - \hat{\beta}_{W} \delta \right)' S^{1/2} P_{b^{\bot}} S^{1/2} \left( \hat{\beta}_{Y} - \hat{\beta}_{W} \delta \right) . \label{eq:cv1}
\end{equation}
The use of (\ref{eq:cv2}) as target function will give the smallest test mean squared error and would be expected to give the more precise estimation. The use of (\ref{eq:cv1}) will tend to select more covariates, since any covariate-outcome effects which are mediated through the risk factor will not be discounted. It will tend to therefore be more conservative in the sense that the standard deviation of the estimates will be larger.

\subsection{Two sample Mendelian randomization}
An advantage of using summarized data is the possibility of using a two sample design for Mendelian randomization. Under this design the genetic variant-risk factor associations and genetic-variant-outcome associations are obtained from separate studies, assumed to be non-overlapping and with similar underlying populations \citep{Hartwig2018}. This allows for a large range combinations of risk factors and outcomes to be considered, since we do not require each trait to have been included in the same study. It also helps to mitigate against the so-called ``winner's curse'' \citep{Taylor2014}, which causes effect estimates to tend to be overestimated in single sample designs.

In the multivariable setting, a two sample approach may in fact involve many samples, with up to one extra sample for each covariate. Again, this is a very flexible design in that it allows for any trait that has been included in published GWAS data to be considered as a potential pleiotropic covariate. It is a valid approach as long as each sample is non-overlapping with the genetic variant-outcome sample and is drawn from a similar underlying population. In practice, these conditions may be somewhat restrictive, particularly in a high-dimensional setting where there are many covariates chosen from a number of GWAS datasets. Some studies are included in the datasets of multiple GWAS consortia, and so there may be overlap with the genetic variant-outcome sample. The extent to which any overlap exists should be checked to ensure it is not substantial. Note that these issues are potential limitations of multivariable Mendelian randomization generally.

\subsection{Inference} \label{se:inference}
Having estimated the causal effect, it is natural to wish to then perform inference, for example, via producing confidence intervals. The post-regularization method will produce a standard error for the causal effect, however the uncertainty is likely to be underestimated since it does not take in to account the model selection event. The fundamental problem is that the same data is being used to both select the covariates to be analysed and to do the analysis itself. A simple and pragmatic approach to get around the problem is to use data splitting, which is a practice that goes (at least as far) back as \cite{Cox1975} (see also, for example, \citealp{Fithian2014}, for further discussion). The idea is to randomly split a dataset into two. One set is used for model selection, the other for inference. The obvious drawback is a loss of power, since the sample size is effectively halved. In our setting, since we are using summarized data, data splitting is not an option. However, using the same logic, we can propose a three sample study design. Here, an independent set of genetic associations is used to perform the regularization method to identify the covariates that should be accounted for. A standard two sample multivariable Mendelian randomization analysis is then performed using separate datasets which contain genetic variant associations with the identified covariates, risk factor and outcome. The independent dataset used for covariate selection should be from a sample which is non-overlapping with those in the analysis datasets and from a similar underlying population.

There is a growing literature on methods for performing inference post model selection without requiring independent samples. \cite{Berk2013} (see also \citealp{Bachoc2019}) propose controlling the family wise error rate across all possible models. In this way, correct coverage of confidence intervals is guaranteed. The same data can be used for both model selection and inference, and furthermore any selection technique can be used, even post-hoc, non-data driven ones. It is, however, very conservative. Furthermore, it is computationally intensive to compute the critical values (the authors note that it begins to be infeasible with $m>20$).

Another strain of literature proposes the selective inference approach \citep{Fithian2014, Lee2016, Tibshirani2016, TaylorTibshirani2018}, where inference is performed conditional on a particular model being chosen. \cite{Lee2016} present a method for computing confidence intervals specifically for the case where the model has been chosen using Lasso. They show that the distribution of the parameter estimators conditional on the model selection event is a truncated normal. The confidence intervals can be very wide when the parameter estimate is close to the boundaries of the truncated normal, which will tend to occur when the signal is weak. It could be expected that this may the case in our Mendelian randomization setting when instruments are typically weak and the number of instruments is moderate. Furthermore, the method is derived for fixed $\lambda$, and so is not valid if the tuning parameter is computed using the data under analysis, for example using cross-validation.

Another approach is to use a double estimation procedure \citep{Belloni2014}. Under this approach, two model selections are performed using standard Lasso. The first selects covariates in the model that regresses $\hat{\beta}_{X}$ on $\hat{\beta}_{W}$. The second selects covariates in the model that regresses $\hat{\beta}_{Y}$ on $\hat{\beta}_{W}$. The set of covariates used in the final model is the union of the two individual sets. The procedure was developed for the scenario where the covariates are determinants of both the risk factor and the outcome. Although this is not the case in the model described in Section 2, in practice there may be associations between the covariates and the risk factor, in which case this method would account for those. In any case, it should provide more conservative confidence intervals than the two sample post-regularization approach.

\section{Simulations} \label{se:sims}

Data on 20,000 individuals were generated from the model given in (\ref{eq:X}), (\ref{eq:Wj}) and (\ref{eq:Y}), with
\begin{align*}
G_{ij} &\sim \text{Binomial} \left( 2, \pi \right), \\
U_{i} &, \:
\varepsilon_{Xi} , \:
\varepsilon_{Wij} , \:
\varepsilon_{Yi} \sim N \left( 0, 1 \right) \: \text{independently}.
\end{align*}
Four scenarios were considered, with different combinations of the number of genetic instruments ($p$) and the number of covariates ($k$): $p=10$, with $k$ either $8$ (scenario 1) or $12$ (scenario 2); and $p=80$, with $k$ either $70$ (scenario 3) or $90$ (scenario 4). We set $\pi=0.3$, $\gamma_{X} = \gamma_{Y} = 1$ and $\gamma_{W1} = \cdots = \gamma_{Wk} = 1/k$. The elements of $\beta_{X}$ were simulated uniformly on the interval $ \left( 0.15, 0.3 \right) $ (scenarios 1 and 2) or $\left( 0.05, 0.12 \right)$ (scenarios 3 and 4). The elements of the $\beta_{Wj}$'s were simulated uniformly on the interval $\left( -0.2, 0.4 \right)$ (scenarios 1 and 2) or $\left( 0.05, 0.12 \right)$ (scenarios 3 and 4). These values give average $R^2$ statistics (that is, the proportion of the variance in the risk factor explained by the genetic variants) of $10.0\%$ (scenarios 1 and 2) and $11.7\%$ (scenarios 3 and 4). The number of covariates representing pleiotropic pathways (that is, the number of $\delta_{j}$'s not equal to zero) was either 1, 2 or 4 in scenarios 1 and 2, and either 7, 21, or 35 in scenarios 3 and 4. The non-zero $\delta_{j}$'s were simulated uniformly on the interval $\left( -0.2, 0.3 \right)$. Note that all instruments in this setting are invalid and the pleiotropy is unbalanced. The causal effect was either $\theta = 0.2$ or $\theta = 0$.

For each scenario and combination of parameters, two independent datasets were generated. In order to produce the summarized data, the genetic variant-risk factor / outcome associations were estimated using simple linear regression on each genetic variant in turn using the first dataset. The estimates of the genetic variant-outcome associations, and their standard errors, were produced in the same way using the second dataset. For each of $1\,000$ replications the causal effect was estimated using the following methods.

\begin{enumerate}
	\item The inverse-variance weighted method (that is, ignoring all covariates) (IVW).
	\item The two step regularization procedure (Reg).
	\item The multivariable inverse-variance weighted method including only the covariates given a non-zero coefficient by the two step regularization procedure (Post-reg).
	\item The multivariable inverse-variance weighted method with all covariates included (MV-All, scenarios 1 and 3 only).
	\item The multivariable inverse-variance weighted method with only truly pleiotropic covariates included (Oracle).
\end{enumerate}

When using the regularization procedure (that is, in methods b and c), the Lasso component of Step 1 was performed using the glmnet package in R \citep{glmnet}. The set of $\lambda$ values used for cross-validation was the set generated by that package with the number of values set at 100. The target function for cross-validation was the mean squared error, given by (\ref{eq:cv2}), and number of folds was $K=10$. In scenarios 2 and 4, when more than $p-2$ covariates were identified, the tuning parameter was increased to the smallest value such that only $p-2$ covariates were selected. The inverse-variance weighted method, and the multivariable inverse-variance weighted methods using the relevant set of covariates (that is, as used in methods c, d, and e), were performed using the MendelianRandomization package in R \citep{Yavorska2017}. The mean and standard deviations of the estimates are shown in Table \ref{tb:sims1}. Figure \ref{fg:mse} (a)--(b) plots the mean squared error for each scenario and method.

In each case, both the Reg and Post-reg estimators are less biased than IVW and have lower standard deviations. The regularized estimators also have lower standard deviations than the full multivariable estimator, and typically performed at least as well in terms of bias. The mean squared error plots show that the regularized estimators, across all scenarios, sit below the IVW and full multivariable estimators and above the oracle estimator.

The simulations described above represent scenarios where each of the genetic variants are associated with each covariate, but where only some of the covariates have an association with the outcome (that is, sparsity in the covariate effects on the outcome). In practice, it will often be that all covariates under consideration are associated with the outcome, but only some of them are associated with the genetic instruments (that is, sparsity in the genetic variant effects on the covariates). In order to demonstrate that our method is an adequate proxy for this situation, the simulations were repeated where all elements of $\delta$ were non-zero and covariates were removed in the true model by setting columns of $\beta_{W}$ to zero. The mean and standard deviations of the estimates from these simulations are shown in Table \ref{tb:sims2}, and the mean squared error for each scenario and method are shown in Figure \ref{fg:mse} (c)--(d). The results are in line with the previous ones.

We next show the results of performing inference using methods discussed in Section \ref{se:inference}. Using the same set of simulations as above (where sparsity is in the covariate effects on the outcome), confidence intervals were computed by performing the multivariable inverse-variance weighted method using sets of covariates which were chosen as follows.

\begin{enumerate}
	\item All covariates ignored (IVW).
	\item The two step regularization procedure using the mean squared error, given by (\ref{eq:cv2}), in cross-validation (2 sample(a)).
	\item The two step regularization procedure where cross-validation was performed independent of the genetic variant-risk factor associations, that is, using (\ref{eq:cv1}) as target function (2 sample(b)).
	\item The two step regularization procedure using an independent sample and the mean squared error, given by (\ref{eq:cv2}), in cross-validation (3 sample(a)).
	\item The two step regularization procedure using an independent sample and where cross-validation was performed independent of the genetic variant-risk factor associations, that is, using (\ref{eq:cv1}) as target function (3 sample(b)).
	\item The double estimation procedure (Double est.).
	\item Only truly pleiotropic covariates included (Oracle).
\end{enumerate}

In each case, the model was fitted using the MendelianRandomization package in R with random effects (that is, allowing over-dispersion, see \citealp{ThompsonSharp1999} and \citealp{BThompson2017interpreting}) and 95\% confidence intervals derived using the normal distribution. The means of the standard errors, coverage (that is, the proportion of confidence intervals containing the true causal effect) and power (that is, the proportion of confidence intervals not containing zero) are shown in Table \ref{tb:sims3} (for $\theta = 0.2$) and Table \ref{tb:sims4} (for $\theta = 0$). Note that in the $\theta=0$ case, power in fact refers to the Type I error rate.

The results show that using the same data to do both covariate selection and inference results in under-coverage, as expected. The three sample approach gives coverage close to the nominal level of 0.95, particularly in scenarios 3 and 4 with larger numbers of instruments. The use of (\ref{eq:cv1}) in cross-validation tends to give coverage closer to 0.95 than the use of (\ref{eq:cv2}), although not uniformly. It also gives wider confidence intervals, suggesting that it is more conservative in covariate selection (that is, tends to give lower levels of sparsity). The double estimation method, while producing better coverage than the two sample approach, in most cases did not reach the 0.95 level. As expected, the IVW method always had the lowest coverage and the full multivariable method (for scenarios 1 and 3) always had the lowest power.

\begin{sidewaystable}
	\centering
	\caption{Mean and standard deviation (SD) of estimates from the various methods when there is sparsity in the covariate effects on the outcome. Scenarios 1 and 2 have either 1, 2 or 4 truly pleiotropic covariates and scenarios 3 and 4 have either 7, 21 or 35 truly pleiotropic covariates.}
	\label{tb:sims1}
	\scalebox{0.9}{
		\begin{tabular}{@{\extracolsep{4pt}}l c c c c c c c c c c c c@{}}
			\hline
			& \multicolumn{6}{c}{$\theta = 0.2$} & \multicolumn{6}{c}{$\theta = 0$} \\ \cline{2-7} \cline{8-13}
			& \multicolumn{2}{c}{1 / 7 Covariates} & \multicolumn{2}{c}{2 / 21 Covariates} & \multicolumn{2}{c}{4 / 35 Covariates} & \multicolumn{2}{c}{1 / 7 Covariates} & \multicolumn{2}{c}{2 / 21 Covariates} & \multicolumn{2}{c}{4 / 35 Covariates}\\ \cline{2-3} \cline{4-5} \cline{6-7} \cline{8-9} \cline{10-11} \cline{12-13}
			Method & Mean & SD & Mean & SD & Mean & SD & Mean & SD & Mean & SD & Mean & SD \\ \hline
			\multicolumn{13}{c}{Scenario 1: $p = 10$, $k = 8$}\\ \hline
			IVW&0.219&0.077&0.240&0.103&0.289&0.146&0.024&0.075&0.040&0.105&0.086&0.142\\
			Reg&0.204&0.060&0.203&0.066&0.217&0.090&0.005&0.053&0.007&0.063&0.012&0.088\\
			Post-reg&0.201&0.066&0.198&0.073&0.210&0.096&0.004&0.059&0.003&0.070&0.006&0.099\\
			MV-All&0.198&0.282&0.188&0.239&0.196&0.252&0.007&0.259&0.000&0.209&-0.015&0.314\\
			Oracle&0.199&0.030&0.198&0.037&0.198&0.058&0.000&0.027&0.001&0.033&0.000&0.048\\ \hline
			\multicolumn{13}{c}{Scenario 2: $p = 10$, $k = 12$}\\ \hline
			IVW&0.222&0.075&0.251&0.107&0.279&0.146&0.022&0.073&0.045&0.105&0.092&0.146\\
			Reg&0.205&0.050&0.212&0.065&0.219&0.095&0.003&0.044&0.008&0.060&0.028&0.089\\
			Post-reg&0.203&0.056&0.208&0.072&0.213&0.104&0.000&0.049&0.005&0.066&0.021&0.093\\
			Oracle&0.198&0.030&0.201&0.038&0.202&0.057&0.000&0.027&0.000&0.032&0.002&0.051\\ \hline
			\multicolumn{13}{c}{Scenario 3: $p = 80$, $k = 70$}\\ \hline
			IVW&0.290&0.110&0.478&0.194&0.678&0.243&0.095&0.110&0.279&0.199&0.480&0.238\\
			Reg&0.200&0.050&0.214&0.080&0.231&0.113&0.014&0.045&0.035&0.079&0.053&0.106\\
			Post-reg&0.181&0.060&0.184&0.088&0.185&0.121&0.003&0.055&0.013&0.084&0.018&0.117\\
			MV-All&0.167&0.178&0.169&0.194&0.160&0.223&-0.005&0.157&0.003&0.192&0.001&0.211\\
			Oracle&0.192&0.032&0.189&0.054&0.180&0.083&0.002&0.029&0.004&0.050&0.006&0.078\\ \hline
			\multicolumn{13}{c}{Scenario 4: $p = 80$, $k = 90$}\\ \hline
			IVW&0.294&0.112&0.485&0.185&0.666&0.249&0.097&0.109&0.291&0.196&0.466&0.245\\
			Reg&0.201&0.053&0.221&0.087&0.255&0.125&0.016&0.045&0.039&0.082&0.072&0.113\\
			Post-reg&0.180&0.066&0.188&0.097&0.211&0.134&0.004&0.055&0.013&0.093&0.035&0.124\\
			Oracle&0.191&0.033&0.187&0.053&0.188&0.080&0.002&0.029&0.003&0.051&0.005&0.073\\ \hline
	\end{tabular}}
\end{sidewaystable}

\begin{sidewaystable}
	\centering
	\caption{Mean and standard deviation (SD) of estimates from the various methods when there is sparsity in the genetic variant effects on the covariates. Scenarios 1 and 2 have either 1, 2 or 4 truly pleiotropic covariates and scenarios 3 and 4 have either 7, 21 or 35 truly pleiotropic covariates.}
	\label{tb:sims2}
	\scalebox{0.9}{
		\begin{tabular}{@{\extracolsep{4pt}}l c c c c c c c c c c c c@{}}
			\hline
			& \multicolumn{6}{c}{$\theta = 0.2$} & \multicolumn{6}{c}{$\theta = 0$} \\ \cline{2-7} \cline{8-13}
			& \multicolumn{2}{c}{1 / 7 Covariates} & \multicolumn{2}{c}{2 / 21 Covariates} & \multicolumn{2}{c}{4 / 35 Covariates} & \multicolumn{2}{c}{1 / 7 Covariates} & \multicolumn{2}{c}{2 / 21 Covariates} & \multicolumn{2}{c}{4 / 35 Covariates}\\ \cline{2-3} \cline{4-5} \cline{6-7} \cline{8-9} \cline{10-11} \cline{12-13}
			Method & Mean & SD & Mean & SD & Mean & SD & Mean & SD & Mean & SD & Mean & SD \\ \hline
			\multicolumn{13}{c}{Scenario 1: $p = 10$, $k = 8$}\\ \hline
			IVW&0.219&0.077&0.240&0.104&0.288&0.146&0.020&0.076&0.040&0.103&0.089&0.146\\
			Reg&0.203&0.046&0.204&0.057&0.215&0.089&0.003&0.044&0.005&0.055&0.013&0.086\\
			Post-reg&0.201&0.047&0.200&0.061&0.207&0.094&0.001&0.044&0.002&0.057&0.005&0.092\\
			MV-All&0.197&0.176&0.199&0.172&0.204&0.199&0.000&0.164&0.000&0.146&0.002&0.171\\
			Oracle&0.199&0.032&0.198&0.039&0.198&0.060&0.000&0.028&-0.001&0.035&-0.001&0.053\\ \hline
			\multicolumn{13}{c}{Scenario 2: $p = 10$, $k = 12$}\\ \hline
			IVW&0.222&0.076&0.251&0.107&0.280&0.147&0.023&0.075&0.051&0.107&0.080&0.146\\
			Reg&0.205&0.044&0.212&0.061&0.223&0.094&0.005&0.040&0.013&0.057&0.023&0.088\\
			Post-reg&0.201&0.045&0.206&0.062&0.217&0.094&0.002&0.040&0.008&0.059&0.018&0.088\\
			Oracle&0.198&0.032&0.200&0.041&0.201&0.060&0.000&0.029&0.002&0.037&0.002&0.054\\ \hline
			\multicolumn{13}{c}{Scenario 3: $p = 80$, $k = 70$}\\ \hline
			IVW&0.290&0.112&0.478&0.194&0.678&0.243&0.096&0.112&0.284&0.193&0.484&0.243\\
			Reg&0.212&0.056&0.237&0.087&0.253&0.118&0.020&0.052&0.047&0.082&0.068&0.112\\
			Post-reg&0.197&0.053&0.208&0.083&0.207&0.116&0.007&0.049&0.022&0.079&0.027&0.112\\
			MV-All&0.188&0.124&0.194&0.168&0.175&0.202&-0.001&0.116&0.011&0.158&-0.002&0.187\\
			Oracle&0.191&0.040&0.189&0.063&0.180&0.092&0.001&0.037&0.006&0.059&0.003&0.086\\ \hline
			\multicolumn{13}{c}{Scenario 4: $p = 80$, $k = 90$}\\ \hline
			IVW&0.294&0.115&0.485&0.186&0.666&0.250&0.100&0.115&0.292&0.186&0.472&0.249\\
			Reg&0.217&0.062&0.253&0.092&0.288&0.131&0.026&0.059&0.063&0.087&0.100&0.126\\
			Post-reg&0.201&0.058&0.219&0.090&0.238&0.127&0.011&0.054&0.033&0.085&0.054&0.121\\
			Oracle&0.191&0.044&0.188&0.063&0.186&0.089&0.002&0.041&0.005&0.059&0.008&0.084\\ \hline
	\end{tabular}}
\end{sidewaystable}

\begin{figure}[hp]
	\centering
	\includegraphics{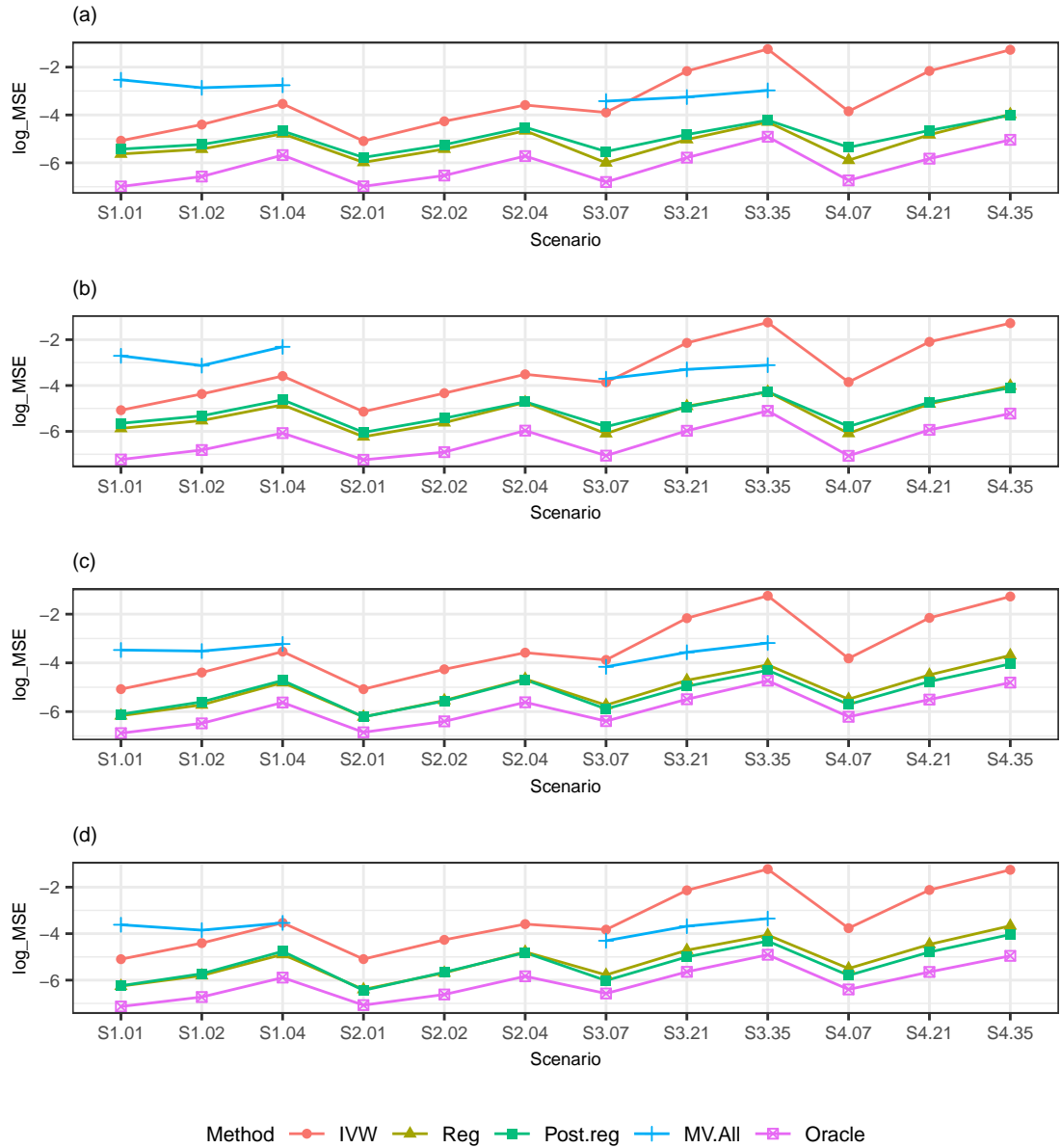}
	\caption{Logarithm of the mean squared errors for each scenario (S1--S4) and number of truly pleiotropic covariates (01--35). Plots (a) and (b), where $\theta=0.2$ and $\theta=0$, respectively, show the results from simulations where there is sparsity in the covariate effects on the outcome. Plots (c) and (d), where $\theta=0.2$ and $\theta=0$, respectively, show the results from simulations where there is sparsity in the genetic variant effects on the covariates.}
	\label{fg:mse}
\end{figure}

\begin{sidewaystable}
	\centering
	\caption{Mean, standard deviation (SD), mean standard errors (SE), coverage (Cov) and power (Pow) of estimates from the various methods (when there is sparsity in the covariate effects on the outcome) with $\theta=0.2$.}
	\label{tb:sims3}
	\scalebox{0.9}{
		\begin{tabular}{@{\extracolsep{4pt}}l c c c c c c c c c c c c c c c@{}}
			\hline
			& \multicolumn{5}{c}{1 / 7 Covariates} & \multicolumn{5}{c}{2 / 21 Covariates} & \multicolumn{5}{c}{4 / 35 Covariates}\\ \cline{2-6} \cline{7-11} \cline{12-16}
			Method & Mean & SD & SE & Cov & Pow & Mean & SD & SE & Cov & Pow & Mean & SD & SE & Cov & Pow \\ \hline
			\multicolumn{16}{c}{Scenario 1: $p = 10$, $k = 8$}\\ \hline
			IVW&0.219&0.077&0.040&0.729&0.958&0.240&0.103&0.054&0.667&0.920&0.289&0.146&0.074&0.595&0.852\\
			2 sample (a)&0.201&0.066&0.050&0.894&0.922&0.198&0.073&0.056&0.888&0.865&0.210&0.096&0.073&0.859&0.788\\
			2 sample (b)&0.201&0.146&0.081&0.900&0.794&0.192&0.143&0.082&0.900&0.757&0.209&0.132&0.090&0.883&0.714\\
			3 sample (a)&0.197&0.074&0.059&0.937&0.891&0.200&0.080&0.065&0.922&0.839&0.208&0.118&0.086&0.914&0.745\\
			3 sample (b)&0.196&0.213&0.097&0.951&0.763&0.194&0.114&0.094&0.939&0.708&0.205&0.135&0.102&0.928&0.666\\
			Double est.&0.202&0.100&0.072&0.926&0.804&0.199&0.132&0.094&0.909&0.736&0.203&0.136&0.104&0.892&0.684\\
			MV-All&0.198&0.282&0.190&0.961&0.425&0.188&0.239&0.196&0.963&0.386&0.196&0.252&0.190&0.957&0.418\\
			Oracle&0.199&0.030&0.032&0.947&0.999&0.198&0.037&0.039&0.956&0.982&0.198&0.058&0.056&0.953&0.903\\ \hline
			\multicolumn{16}{c}{Scenario 2: $p = 10$, $k = 12$}\\ \hline
			IVW&0.222&0.075&0.041&0.726&0.967&0.251&0.107&0.055&0.658&0.928&0.279&0.146&0.074&0.607&0.826\\
			2 sample (a)&0.203&0.056&0.049&0.910&0.911&0.208&0.072&0.060&0.878&0.855&0.213&0.104&0.072&0.810&0.789\\
			2 sample (b)&0.201&0.087&0.082&0.948&0.706&0.204&0.089&0.085&0.923&0.703&0.213&0.114&0.090&0.868&0.695\\
			3 sample (a)&0.201&0.057&0.053&0.940&0.908&0.205&0.086&0.071&0.930&0.811&0.207&0.118&0.090&0.882&0.702\\
			3 sample (b)&0.200&0.098&0.096&0.962&0.673&0.203&0.111&0.100&0.948&0.633&0.200&0.131&0.113&0.928&0.581\\
			Double est.&0.201&0.072&0.062&0.935&0.850&0.199&0.079&0.075&0.906&0.779&0.209&0.111&0.084&0.826&0.723\\
			Oracle&0.198&0.030&0.032&0.955&0.997&0.201&0.038&0.040&0.956&0.986&0.202&0.057&0.057&0.957&0.909\\ \hline
			\multicolumn{16}{c}{Scenario 3: $p = 80$, $k = 70$}\\ \hline
			IVW&0.290&0.110&0.043&0.392&0.970&0.478&0.194&0.069&0.216&0.975&0.678&0.243&0.089&0.081&0.988\\
			2 sample (a)&0.181&0.060&0.044&0.846&0.923&0.184&0.088&0.069&0.879&0.719&0.185&0.121&0.091&0.848&0.539\\
			2 sample (b)&0.180&0.062&0.045&0.836&0.908&0.185&0.091&0.070&0.867&0.717&0.185&0.122&0.093&0.853&0.529\\
			3 sample (a)&0.186&0.051&0.049&0.947&0.936&0.186&0.083&0.084&0.951&0.618&0.186&0.120&0.120&0.941&0.370\\
			3 sample (b)&0.187&0.051&0.050&0.943&0.938&0.187&0.084&0.084&0.960&0.623&0.186&0.122&0.122&0.931&0.357\\
			Double est.&0.185&0.058&0.048&0.871&0.918&0.185&0.084&0.073&0.911&0.720&0.191&0.114&0.096&0.895&0.522\\
			MV-All&0.167&0.178&0.181&0.947&0.153&0.169&0.194&0.200&0.952&0.145&0.160&0.223&0.223&0.943&0.131\\
			Oracle&0.192&0.032&0.033&0.950&1.000&0.189&0.054&0.054&0.942&0.941&0.180&0.083&0.080&0.939&0.615\\ \hline
			\multicolumn{16}{c}{Scenario 4: $p = 80$, $k = 90$}\\ \hline
			IVW&0.294&0.112&0.043&0.393&0.977&0.485&0.185&0.070&0.198&0.981&0.666&0.249&0.088&0.099&0.982\\
			2 sample (a)&0.180&0.066&0.046&0.802&0.895&0.188&0.097&0.074&0.859&0.686&0.211&0.134&0.099&0.837&0.574\\
			2 sample (b)&0.179&0.066&0.047&0.803&0.897&0.190&0.099&0.076&0.856&0.669&0.210&0.136&0.103&0.845&0.544\\
			3 sample (a)&0.187&0.055&0.052&0.930&0.908&0.192&0.097&0.093&0.936&0.576&0.202&0.153&0.142&0.940&0.368\\
			3 sample (b)&0.185&0.056&0.054&0.936&0.883&0.189&0.101&0.096&0.938&0.564&0.201&0.152&0.143&0.946&0.375\\
			Double est.&0.185&0.063&0.050&0.874&0.884&0.188&0.093&0.080&0.912&0.648&0.206&0.134&0.113&0.888&0.521\\
			Oracle&0.191&0.033&0.033&0.938&1.000&0.187&0.053&0.054&0.955&0.935&0.188&0.080&0.079&0.950&0.660\\ \hline
	\end{tabular}}
\end{sidewaystable}

\begin{sidewaystable}
	\centering
	\caption{Mean, standard deviation (SD), mean standard errors (SE), coverage (Cov) and power (Pow) of estimates from the various methods (when there is sparsity in the covariate effects on the outcome) with $\theta=0$.}
	\label{tb:sims4}
	\scalebox{0.9}{
		\begin{tabular}{@{\extracolsep{4pt}}l c c c c c c c c c c c c c c c@{}}
			\hline
			& \multicolumn{5}{c}{1 / 7 Covariates} & \multicolumn{5}{c}{2 / 21 Covariates} & \multicolumn{5}{c}{4 / 35 Covariates}\\ \cline{2-6} \cline{7-11} \cline{12-16}
			Method & Mean & SD & SE & Cov & Pow & Mean & SD & SE & Cov & Pow & Mean & SD & SE & Cov & Pow \\ \hline
			\multicolumn{16}{c}{Scenario 1: $p = 10$, $k = 8$}\\ \hline
			IVW&0.024&0.075&0.040&0.707&0.293&0.040&0.105&0.053&0.647&0.353&0.086&0.142&0.073&0.587&0.413\\
			2 sample (a)&0.004&0.059&0.044&0.905&0.095&0.003&0.070&0.053&0.901&0.099&0.006&0.099&0.067&0.863&0.137\\
			2 sample (b)&0.000&0.171&0.072&0.924&0.076&0.000&0.125&0.075&0.906&0.094&0.001&0.138&0.081&0.874&0.126\\
			3 sample (a)&-0.003&0.100&0.051&0.936&0.064&0.002&0.065&0.060&0.934&0.066&0.000&0.089&0.074&0.904&0.096\\
			3 sample (b)&-0.002&0.148&0.086&0.950&0.050&0.007&0.105&0.079&0.942&0.058&0.001&0.110&0.091&0.934&0.066\\
			Double est.&0.003&0.126&0.072&0.935&0.065&0.003&0.109&0.081&0.929&0.071&-0.007&0.277&0.104&0.916&0.084\\
			MV-All&0.007&0.259&0.173&0.964&0.036&0.000&0.209&0.163&0.956&0.044&-0.015&0.314&0.170&0.955&0.045\\
			Oracle&0.000&0.027&0.028&0.964&0.036&0.001&0.033&0.035&0.954&0.046&0.000&0.048&0.050&0.965&0.035\\ \hline
			\multicolumn{16}{c}{Scenario 2: $p = 10$, $k = 12$}\\ \hline
			IVW&0.022&0.073&0.039&0.746&0.254&0.045&0.105&0.054&0.653&0.347&0.092&0.146&0.074&0.556&0.444\\
			2 sample (a)&0.000&0.049&0.045&0.916&0.084&0.005&0.066&0.053&0.886&0.114&0.021&0.093&0.065&0.815&0.185\\
			2 sample (b)&-0.001&0.071&0.072&0.941&0.059&0.005&0.081&0.074&0.928&0.072&0.016&0.093&0.078&0.881&0.119\\
			3 sample (a)&-0.001&0.053&0.050&0.944&0.056&0.005&0.073&0.060&0.907&0.093&0.017&0.103&0.078&0.860&0.140\\
			3 sample (b)&-0.005&0.086&0.089&0.960&0.040&-0.002&0.099&0.085&0.945&0.055&0.008&0.121&0.101&0.919&0.081\\
			Double est.&-0.002&0.059&0.057&0.933&0.067&0.003&0.073&0.066&0.922&0.078&0.025&0.108&0.080&0.838&0.162\\
			Oracle&0.000&0.027&0.029&0.967&0.033&0.000&0.032&0.035&0.969&0.031&0.002&0.051&0.051&0.968&0.032\\ \hline
			\multicolumn{16}{c}{Scenario 3: $p = 80$, $k = 70$}\\ \hline
			IVW&0.095&0.110&0.042&0.376&0.624&0.279&0.199&0.067&0.209&0.791&0.480&0.238&0.087&0.086&0.914\\
			2 sample (a)&0.003&0.055&0.039&0.845&0.155&0.013&0.084&0.065&0.867&0.133&0.018&0.117&0.087&0.859&0.141\\
			2 sample (b)&0.002&0.055&0.040&0.846&0.154&0.013&0.086&0.066&0.871&0.129&0.018&0.122&0.089&0.858&0.142\\
			3 sample (a)&0.005&0.046&0.044&0.931&0.069&0.010&0.083&0.077&0.939&0.061&0.016&0.123&0.112&0.914&0.086\\
			3 sample (b)&0.005&0.048&0.045&0.934&0.066&0.013&0.083&0.078&0.942&0.058&0.018&0.119&0.113&0.919&0.081\\
			Double est.&0.006&0.054&0.043&0.865&0.135&0.020&0.080&0.069&0.886&0.114&0.028&0.109&0.092&0.884&0.116\\
			MV-All&-0.005&0.157&0.161&0.957&0.043&0.003&0.192&0.184&0.946&0.054&0.001&0.211&0.206&0.947&0.053\\
			Oracle&0.002&0.029&0.030&0.947&0.053&0.004&0.050&0.049&0.943&0.057&0.006&0.078&0.073&0.928&0.072\\ \hline
			\multicolumn{16}{c}{Scenario 4: $p = 80$, $k = 90$}\\ \hline
			IVW&0.097&0.109&0.042&0.381&0.619&0.291&0.196&0.068&0.189&0.811&0.466&0.245&0.087&0.110&0.890\\
			2 sample (a)&0.004&0.055&0.041&0.850&0.150&0.013&0.093&0.070&0.847&0.153&0.035&0.124&0.096&0.854&0.146\\
			2 sample (b)&0.005&0.055&0.041&0.857&0.143&0.011&0.096&0.072&0.851&0.149&0.035&0.129&0.100&0.858&0.142\\
			3 sample (a)&0.005&0.047&0.046&0.942&0.058&0.011&0.092&0.085&0.923&0.077&0.019&0.139&0.133&0.945&0.055\\
			3 sample (b)&0.005&0.049&0.047&0.944&0.056&0.011&0.096&0.088&0.933&0.067&0.017&0.144&0.134&0.946&0.054\\
			Double est.&0.008&0.055&0.046&0.884&0.116&0.021&0.088&0.076&0.889&0.111&0.048&0.119&0.108&0.871&0.129\\
			Oracle&0.002&0.029&0.030&0.956&0.044&0.003&0.051&0.049&0.940&0.060&0.005&0.073&0.074&0.950&0.050\\ \hline
	\end{tabular}}
\end{sidewaystable}

\section{Investigating the causal effect of urate concentration on coronary heart disease} \label{se:appliedex}
We consider the study of \cite{White2016} looking at the effect of plasma urate concentration on coronary heart disease. The study identified 31 genetic variants associated with urate concentration at a genome-wide significance level of $5 \times 10^{-8}$. Summarized associations between these genetic variants and urate concentration were produced from a combination of published meta-analyses. The associations between the genetic variants and coronary heart disease were obtained from the CARDIoGRAMplusC4D study. In addition, eight potential pleiotropic covariates were identified: Fasting glucose, BMI, type 2 diabetes, HDL cholesterol, LDL cholesterol, triglycerides, systolic blood pressure and diastolic blood pressure. These covariates were chosen as risk factors which have been shown observationally to be associated with increased urate concentration, and are also known risk factors for coronary heart disease. By examining the associations between the 31 genetic variants and the covariates, \cite{White2016} concluded that four of them were potential sources of pleiotropy: HDL cholesterol, triglycerides, systolic blood pressure and diastolic blood pressure. A Mendelian randomization analysis ignoring covariates suggests that urate concentration has a causal effect on coronary heart disease. However, when including the covariates in the model, the results suggest that there is no causal effect. This is supported by \cite{Bowden20172sample}, who analysed the same data using the MR-Egger method.

We re-analysed the causal effect of urate concentration on coronary heart disease using our regularization procedure. Details of each of the data sources for the genetic variant associations are given in the Appendix (noting there are some differences in the data sources used here to those used by \citealp{White2016}). Figure \ref{fg:delta_lam} shows the values of the coefficients for the genetic variant-risk factor and genetic-variant-covariate associations produced by the regularization procedure for increasing values of $\lambda$. The value of $\lambda$ used in the final model was chosen by performing 10-fold cross-validation $100$ times and taking the mean minimizer of the mean squared error. This value is indicated in Figure \ref{fg:delta_lam} by the vertical dashed line. The procedure identified two covariates that should be included in the analysis: diastolic blood pressure and BMI. This suggests that pleiotropy is being caused by these two covariates only. Interestingly, BMI was not identified by \cite{White2016} as a covariate to be included in the model. Furthermore, HDL cholesterol was the final covariate to be included using the regularization method, whereas it was one of the four chosen by \cite{White2016}.

We performed multivariable Mendelian randomization analyses using five sets of covariates: no covariates; all covariates; the four covariates identified by \cite{White2016}; diastolic blood pressure only; and diastolic blood pressure and BMI. Table \ref{tb:urate} shows the estimates of the log causal odds ratio for each model, as well as their standard errors and 95\% confidence intervals (computed using a random effects model and the normal distribution). In agreement with the previous studies, the results suggest that urate concentration has a causal effect on coronary heart disease when ignoring covariates. When covariates are included, the results suggest that there is no causal effect. The causal effect estimate when only diastolic blood pressure (0.036) or diastolic blood pressure and BMI (0.034) were included are close to the estimates obtained by including all covariates (0.036) or the set of covariates chosen by \cite{White2016} (0.038).

We use a covariate balancing plot which shows the correlation between the genetic variant-risk factor / covariate associations and the residuals obtained after regressing the genetic variant-outcome associations on the genetic variant-covariate associations for each set of covariates considered. If there is no pleiotropy exerted by the covariates, or there is pleiotropy but the model has accounted for it, the correlations with the genetic variant-covariate associations will be close to zero. The plot thus demonstrates two things: the strength of the association between the instruments and the risk factor when controlling for the different sets of covariates (shown by the size of the correlation with urate concentration), and how well each model has balanced the pleiotropic effects (shown by the size of the correlations with the covariates). Figure \ref{fg:balance} shows that, when all covariates are ignored, the genetic variant-risk factor correlation is the strongest, but there are also strong correlations with all other covariates except for glucose fasting. When all covariates are included, the genetic variant-risk factor strength is somewhat weaker, but all covariates are close to uncorrelated with the residuals. When the four covariates of \cite{White2016} are included, the covariates are reasonably balanced except for BMI, and a similar pattern is seen when diastolic blood pressure only is included. When diastolic blood pressure and BMI are included, a similar pattern of instrument-exposure correlation and covariate balance is seen as in the case of the full multivariable model.

It should be noted that the analysis here has been performed in a two-sample framework, implying that the confidence interval from the model chosen by the regularization procedure could be too narrow. This would be the case if further covariates should be included to account for pleiotropy. However, the covariate balancing plot suggests that the inclusion of more covariates is not needed. Furthermore, since we have a finding of no causal effect, widening the confidence interval will only strengthen the evidence behind this finding.

\begin{table}
	\centering
	\caption{Estimates, standard errors (SE) and 95\% confidence intervals of the log causal odds ratio for coronary heart disease per one standard deviation increase in plasma urate concentration levels.}
	\label{tb:urate}
	\begin{tabular}{l c c c}
		\hline
		Covariates included & Estimate & SE & 95\% Confidence interval\\ \hline
		None& 0.104&0.040&(0.025, 0.182)\\
		All& 0.036&0.031&(-0.025, 0.096)\\
		HDL, Tri, SBP \& DBP & 0.038&0.029&(-0.020, 0.095)\\
		DBP&0.036&0.027&(-0.017, 0.089)\\
		DBP \& BMI&0.034&0.027&(-0.019, 0.087)\\ \hline
	\end{tabular}
\end{table}

\begin{figure}
	\centering
	\includegraphics{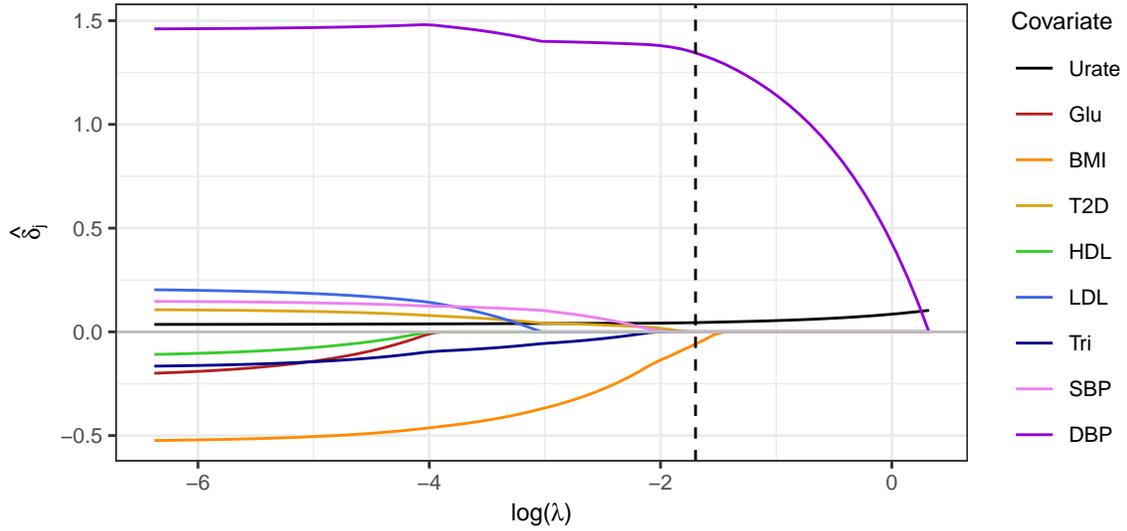}
	\caption{Estimates of regression coefficients for the genetic variant-risk factor association and the eight genetic variant-covariate associations for different values of $\lambda$. The dashed vertical line indicates the value of $\lambda$ chosen by cross-validation.}
	\label{fg:delta_lam}
\end{figure}

\begin{figure}
	\centering
	\includegraphics{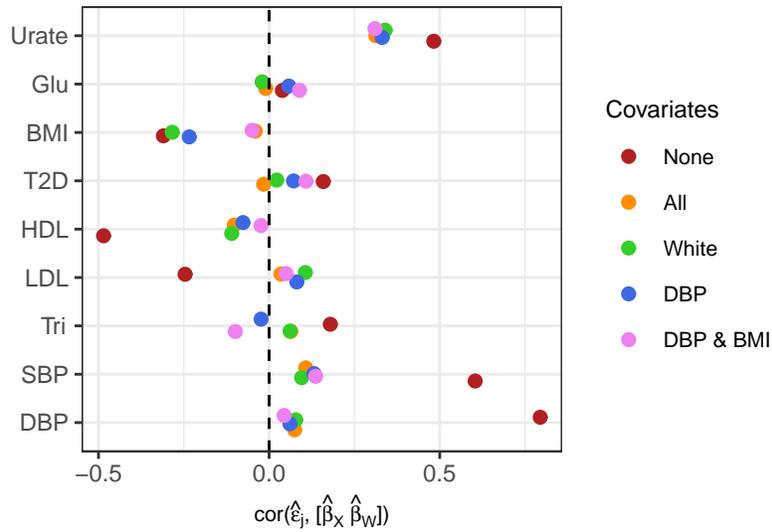}
	\caption{Correlation between the genetic variant-risk factor association and genetic variant-covariate associations, and the residuals obtained after regressing the genetic variant-outcome association on each set of genetic variant-covariate associations.}
	\label{fg:balance}
\end{figure}

\section{Discussion}
In this paper, we have presented a method for estimating a causal effect of a risk factor on an outcome in a Mendelian randomization setting in the face of pleiotropy. The method does not require any of the genetic variants to be valid instruments and can be performed using summarized data only. By controlling for covariates that have pathways to the outcome which bypass the risk factor we remove the bias that arises due to these covariates. By not controlling for further covariates unnecessarily, we gain a more precise estimate than we would from a full multivariable model. Furthermore, we can consider cases where there are more covariates than instruments, which cannot be handled by standard multivariable Mendelian randomization techniques. We also discussed different ways of constructing confidence intervals for the causal effect. Simulations suggest that our proposed three sample approach produces valid confidence intervals and can be used to infer the presence of a causal effect.

The regularization method can be used as a model selection tool which identifies mechanisms by which sets of genetic variants influence an outcome. In the study looking at the effect of urate concentration on coronary heart disease, the method suggested that pleiotropy was being caused by diastolic blood pressure and BMI. This is in contrast to the previous finding suggesting that HDL cholesterol, triglycerides and systolic blood pressure, but not BMI, should be accounted for.

Although the method was derived for the case where there are no causal pathways between the risk factor and covariates, the method will still work when such pathways exist. In practice, there may be many pathways among the covariates if the set of potential covariates is chosen by taking all plausible traits from a GWAS database, many of which may be highly correlated. Two possible scenarios relating to causal pathways between the risk factor and the covariates are illustrated via the directed acyclic graphs in Figure \ref{fg:dag2}. The first is where covariates are predictors of both the risk factor and the outcome. The second is where covariates are mediators of the causal effect of the risk factor on the outcome (also known as vertical pleiotropy). In the latter case, care needs to be taken over the interpretation of the causal effect estimate. The estimate in this case will be of the direct causal effect of the risk factor on the outcome, and will not include the indirect effect via the mediator(s) \citep{Burgess2017mediator,Sanderson2018}. Hence an alternative interpretation of the results of the applied example is that the covariates (in particular, diastolic blood pressure) are mediators of the effect of urate concentration on coronary heart disease.

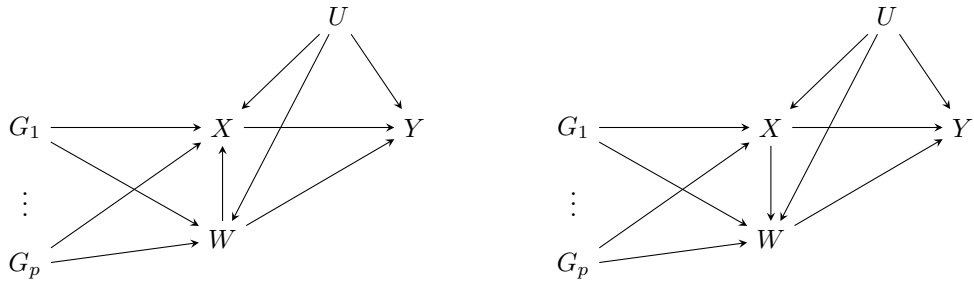
\begin{figure}
	\centering
	\begin{tikzpicture}[]
	\node[] (X) {$X$};
	\node[] (G1) [left=2cm of X] {$G_{1}$};
	\node[] (Gdots) [below=0.25cm of G1] {$\vdots$};
	\node[] (Gp) [below=0.25cm of Gdots] {$G_{p}$};
	\node[] (W) [below=of X] {$W$};
	\node[] (U) [above right=1.41cm of X] {$U$};
	\node[] (Y) [right=2cm of X] {$Y$};
	
	\path[-stealth]
	(G1) edge (X)
	(G1) edge (W)
	(Gp) edge (X)
	(Gp) edge (W)
	(W) edge (Y)
	(X) edge (Y)
	(U) edge (X)
	(U) edge (Y)
	(U) edge (W)
	(W) edge (X);
	\end{tikzpicture}
	\quad\quad\quad\quad
	\begin{tikzpicture}[]
	\node[] (X) {$X$};
	\node[] (G1) [left=2cm of X] {$G_{1}$};
	\node[] (Gdots) [below=0.25cm of G1] {$\vdots$};
	\node[] (Gp) [below=0.25cm of Gdots] {$G_{p}$};
	\node[] (W) [below=of X] {$W$};
	\node[] (U) [above right=1.41cm of X] {$U$};
	\node[] (Y) [right=2cm of X] {$Y$};
	
	\path[-stealth]
	(G1) edge (X)
	(G1) edge (W)
	(Gp) edge (X)
	(Gp) edge (W)
	(W) edge (Y)
	(X) edge (Y)
	(U) edge (X)
	(U) edge (Y)
	(U) edge (W)
	(X) edge (W);
	\end{tikzpicture}
	\caption{Directed acyclic graphs showing the cases where $W$ has a causal effect on both $X$ and $Y$ (left graph) and when $W$ is a mediator of the causal effect of $X$ on $Y$ (right graph). Here, $W$ represents one or more, possibly correlated, covariates.}
	\label{fg:dag2}
\end{figure}

We note that we are assuming linearity in the relationships between the genetic variants and the risk factor, covariates and the outcome. Although we have derived the method as though we have a continuous outcome, an advantage of using summarized data is that it allows us to also consider binary outcomes. In this case, the $\hat{\beta}_{Yj}$'s represent estimates of log odds ratios obtained by fitting logistic regression models. Note, however, that, due to the non-collapsibility of the odds ratio, causal effect estimates will tend toward the null \citep{Vansteelandt2011, Bowden20172sample}.

In summary, the method we have developed provides a causal effect estimator in a Mendelian randomization setting where potentially all genetic instruments are invalid due to pleiotropy via a possibly high-dimensional set of covariates. The estimator is robust to measured pleiotropy and more efficient than existing methods.

\section{Software} \label{se:software}

R code for performing the proposed method and for generating the simulation results are available at \url{https://github.com/aj-grant/mvcovreg}.

\section*{Acknowledgments}

The authors thank Stijn Vansteelandt and Verena Zuber for helpful discussions in the development of this work.

\section*{Funding}
Andrew J. Grant and Stephen Burgess are supported by a Sir Henry Dale Fellowship jointly funded by the Wellcome Trust and the Royal Society (grant number 204623/Z/16/Z). This research was funded by the NIHR Cambridge Biomedical Research Centre. The views expressed are those of the authors and not necessarily those of the NHS, the NIHR or the Department of Health and Social Care.

\appendix

\section{Appendix}

\subsection{Equivalence of covariate balancing allele scores and the multivariable inverse-variance weighted method}

Let $B = \begin{bmatrix} \hat{\beta}_{X} &  \hat{\beta}_{W} \end{bmatrix}$. The multivariable inverse-weighted estimator for $\begin{bmatrix} \theta & \delta' \end{bmatrix}'$ is obtained by regressing $S^{1/2} \hat{\beta}_{Y}$ on $S^{1/2} B$, that is,
\[
\left( B' S B \right)^{-1} \left( B' S \hat{\beta}_{Y} \right)
= \begin{bmatrix} \hat{\beta}_{X}' S \hat{\beta}_{X} & \hat{\beta}_{X}' S \hat{\beta}_{W} \\ \hat{\beta}_{W}' S \hat{\beta}_{X} & \hat{\beta}_{W}' S \hat{\beta}_{W} \end{bmatrix}^{-1} \begin{bmatrix} \hat{\beta}_{X}' S \hat{\beta}_{Y} \\ \hat{\beta}_{W}' S \hat{\beta}_{Y} \end{bmatrix}.
\]
The estimator of $\theta$ is the first entry of this vector. By the matrix inversion lemma, the top row of the first term on the right hand side is
\[
\begin{bmatrix} \frac{1}{V} & -\frac{1}{V} \left( \hat{\beta}_{X}' S \hat{\beta}_{W} \right) \left( \hat{\beta}_{W}' S \hat{\beta}_{W} \right)^{-1} \end{bmatrix}
\]
where
\[
V = \hat{\beta}_{X}' S \hat{\beta}_{X} - \left( \hat{\beta}_{X}' S \hat{\beta}_{W} \right) \left( \hat{\beta}_{W}' S \hat{\beta}_{W} \right)^{-1} \left( \hat{\beta}_{W}' S \hat{\beta}_{X} \right) .
\]
The estimator for $\theta$ is thus
\begin{align*}
&\frac{\hat{\beta}_{X}' S \hat{\beta}_{Y} - \left( \hat{\beta}_{X}' S \hat{\beta}_{W} \right) \left( \hat{\beta}_{W}' S \hat{\beta}_{W} \right)^{-1} \left( \hat{\beta}_{W}' S \hat{\beta}_{Y} \right)}{\hat{\beta}_{X}' S \hat{\beta}_{X} - \left( \hat{\beta}_{X}' S \hat{\beta}_{W} \right) \left( \hat{\beta}_{W}' S \hat{\beta}_{W} \right)^{-1} \left( \hat{\beta}_{W}' S \hat{\beta}_{X} \right)} \\
&=\frac{\left\lbrace \hat{\beta}_{X} - \hat{\beta}_{W} \left( \hat{\beta}_{W}' S \hat{\beta}_{W} \right)^{-1}  \left( \hat{\beta}_{W}' S \hat{\beta}_{X} \right) \right\rbrace' S \hat{\beta}_{Y}}{\left\lbrace \hat{\beta}_{X} - \hat{\beta}_{W} \left( \hat{\beta}_{W}' S \hat{\beta}_{W} \right)^{-1}  \left( \hat{\beta}_{W}' S \hat{\beta}_{X} \right) \right\rbrace' S \hat{\beta}_{X}} \\
&=\frac{\tilde{\alpha}' S \hat{\beta}_{Y}}{\tilde{\alpha}' S \hat{\beta}_{X}} .
\end{align*}

\subsection{Derivation of the two step estimation procedure}

We wish to find
\begin{equation}
\argmin_{\theta, \delta} \frac{1}{2}\left( \hat{\beta}_{Y} - \theta \hat{\beta}_{X} -  \hat{\beta}_{W} \delta \right)' S \left( \hat{\beta}_{Y} - \theta \hat{\beta}_{X} -  \hat{\beta}_{W} \delta \right) + \lambda \sum_{i=1}^{k} \left| \delta_{i} \right|. \label{eq:kreg}
\end{equation}
Following the notation of \cite{Kang2016}, we let $P_{M} = M \left( M' M \right)^{-1} M'$ for some matrix $M$ with $d$ rows such that $M' M$ is invertible, and $P_{M^{\bot}}=I_{d} - P_{M}$. Note that $P_{M} P_{M} = P_{M^{\bot}} P_{M^{\bot}} = P_{M}$, $P_{M} P_{M^{\bot}} = 0$, $P_{M} + P_{M^{\bot}} = I_{d}$ and $P_{M} M = M$. Denoting by $\left\lVert \cdot \right\rVert_{2}$ the $\ell_{2}$ norm, (\ref{eq:kreg}) can be written as
\begin{equation}
\frac{1}{2}\argmin_{\theta, \delta} \left\lVert S^{1/2} \left( \hat{\beta}_{Y} - \theta \hat{\beta}_{X} -  \hat{\beta}_{W} \delta \right) \right\rVert_{2}^{2} + \lambda \sum_{i=1}^{k} \left| \delta_{i} \right|. \label{eq:sspen}
\end{equation}
Let $b = S^{1/2} \hat{\beta}_{X}$. Then
\begin{align}
&\frac{1}{2}\left\lVert S^{1/2} \left( \hat{\beta}_{Y} - \theta \hat{\beta}_{X} -  \hat{\beta}_{W} \delta \right) \right\rVert_{2}^{2} + \lambda \sum_{i=1}^{k} \left| \delta_{i} \right| \nonumber \\
&= \frac{1}{2} \left\lVert \left( P_{b} + P_{b^{\bot}} \right) S^{1/2} \left( \hat{\beta}_{Y} - \theta \hat{\beta}_{X} -  \hat{\beta}_{W} \delta \right) \right\rVert_{2}^{2} + \lambda \sum_{i=1}^{k} \left| \delta_{i} \right| \nonumber \\
&= \frac{1}{2} \left\lVert P_{b} S^{1/2} \left( \hat{\beta}_{Y} - \theta \hat{\beta}_{X} - \hat{\beta}_{W} \delta \right) \right\rVert_{2}^{2} + \frac{1}{2} \left\lVert  P_{b^{\bot}} S^{1/2} \left( \hat{\beta}_{Y} - \theta \hat{\beta}_{X} - \hat{\beta}_{W} \delta \right) \right\rVert_{2}^{2} + \lambda \sum_{i=1}^{k} \left| \delta_{i} \right| \nonumber \\
&= \frac{1}{2} \left\lVert P_{b} S^{1/2} \left( \hat{\beta}_{Y} - \hat{\beta}_{W} \delta \right) - \theta S^{1/2} \hat{\beta}_{X} \right\rVert_{2}^{2} + \frac{1}{2} \left\lVert  P_{b^{\bot}} S^{1/2} \hat{\beta}_{Y} - P_{b^{\bot}} S^{1/2} \hat{\beta}_{W} \delta \right\rVert_{2}^{2} + \lambda \sum_{i=1}^{k} \left| \delta_{i} \right| . \label{eq:3term}
\end{align}
The second and third terms of (\ref{eq:3term}) are independent of $\theta$. The first term of (\ref{eq:3term}) can be set to zero for any value of $\delta = \delta^{*}$ by putting
\begin{equation}
\theta = \frac{\left( \hat{\beta}_{Y} - \hat{\beta}_{W} \delta^{*} \right)'S\hat{\beta}_{X}}{\hat{\beta}_{X}'S\hat{\beta}_{X}} . \label{eq:thmin}
\end{equation}
Thus, (\ref{eq:sspen}) can be solved by minimising the second and third terms of (\ref{eq:3term}) with respect to $\delta$, then setting $\theta$ according to (\ref{eq:thmin}). This is the two step procedure.

\subsection{Data sources for the applied analysis}

The associations between the genetic variants and urate concentration were taken from \cite{White2016}. Note that, although the singificance level for inclusion of a genetic variant was $5 \times 10^{-8}$, one variant (rs164009) which had a p-value larger than $5 \times 10^{-8}$, and less than $5 \times 10^{-7}$, was also included on the basis of a known biological role in urate metabolism. The associations between the genetic variants and coronary heart disease as well as the covariates were taken from GWAS data as summarized in Table \ref{tb:gwas}, and accessed using PhenoScanner \citep{Phenoscanner1, Phenoscanner2}.

\begin{table}[ht]
	\centering
	\caption{Sources of associations between the 31 genetic variants and the covariates.}
	\label{tb:gwas}
	\begin{tabular}{l l l r}
		\hline
		Trait & Consortium & Study & Sample Size\\ \hline
		Coronary heart disease & CARDIoGRAMplusC4D & \cite{cardiogram} & 184\,305 \\
		Fasting glucose & MAGIC & \cite{magic} & 46\,186\\
		BMI & GIANT & \cite{giant} & 339,224\\
		Type 2 diabetes & DIAGRAM & \cite{diagram} & 159\,208\\
		HDL cholesterol & GLGC & \cite{glgc} & 187\,167\\
		LDL cholesterol & GLGC & \cite{glgc} & 173\,082\\
		Triglycerides & GLGC & \cite{glgc} & 177\,861\\
		Systolic blood pressure & Neale Lab & 2017 results & 337\,199\\
		Diastolic blood pressure & Neale Lab & 2017 results & 337\,199\\ \hline
	\end{tabular}
\end{table}

Note that the analysis by \cite{White2016} used the 2013 CARDIoGRAMplusC4D dataset, whereas here we use the 2015 dataset. Similarly, \cite{White2016} used the 2012 dataset from DIAGRAM and the 2010 dataset from GIANT, whereas here we use the 2017 and 2015 datasets, respectively. Finally, \cite{White2016} obtained genetic variant associations with the blood pressure traits from the ICBP consortium, whereas here we use the 2017 results from the analysis of UK Biobank by the Neale Lab 
(http://www.nealelab.is/blog/2017/7/19/rapid-gwas-of-thousands-of-phenotypes-for-337000-samples-in-the-uk-biobank).

\bibliographystyle{apalike}
\bibliography{covariate_balancing_mr_bib}

\end{document}